\documentclass[sigconf]{acmart}

\def\BibTeX{{\rm B\kern-.05em{\sc i\kern-.025em b}\kern-.08emT\kern-.1667em\lower.7ex\hbox{E}\kern-.125emX}}
    
\copyrightyear{2019}
\acmYear{2019}
\acmConference[Technical Report]{Technical Report 2019}{2019}{}
\setcopyright{rightsretained}

\begin{document}

%
\title{Experimental Augmented Reality User Experience}

%

\author{Josef Spjut}
\affiliation{NVIDIA}

\author{Fengyuan Zhu}
\affiliation{NVIDIA}
\affiliation{University of Toronto}

\author{Xiaolei Huang}
\affiliation{NVIDIA}
\affiliation{New York University}

\author{Yichen Shou}
\affiliation{NVIDIA}
\affiliation{University of Pennsylvania}

\author{Ben Boudaoud}

\author{Omer Shapira}

\author{Morgan McGuire}
\affiliation{NVIDIA}


%
\renewcommand{\shortauthors}{Spjut, et al.}

%
\begin{abstract}
Augmented Reality (AR) is an emerging field ripe for experimentation, especially when it comes to developing the kinds of applications and experiences that will drive mass adoption of the technology. While we aren't aware of any current consumer product that realize a wearable, wide Field of View (FoV), AR Head Mounted Display (HMD), such devices will certainly come. In order for these sophisticated, likely high-cost hardware products to succeed, it is important they provide a high quality user experience. To that end, we prototyped 4 experimental applications for wide FoV displays that will likely exist in the future. Given current AR HMD limitations, we used a AR simulator built on web technology and VR headsets to demonstrate these applications, allowing users and designers to peer into the future.
\end{abstract}

%
%


%

%
\begin{teaserfigure}
  \centering
  \includegraphics[height=1.45in]{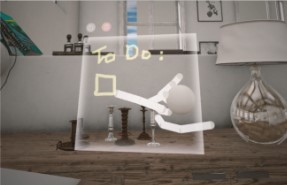}
  \includegraphics[height=1.45in]{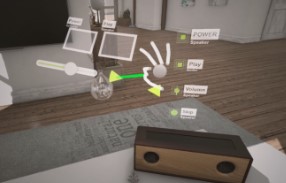}
  \includegraphics[height=1.45in]{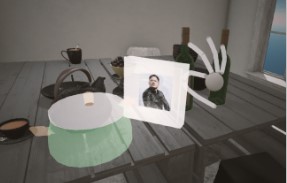}
  \caption{Example AR Applications. From left to right, sticky notes, smart home control and online shopping preview.}
  \Description{Sample AR applications}
  \label{fig:teaser}
\end{teaserfigure}

%
\maketitle

\section{Overview}
We refer to Augmented Reality (AR) as combined real-world (physical) and virtual (rendered) content. This work focuses  on AR Head Mounted Displays (HMDs), and more specifically non-occlusive, optical pass-through designs, which can only add light to that of the real-world "scene" as opposed to completely reproducing scenes through a camera/display (often referred to as "video pass-through"). 
Current optical pass-through AR HMD products like Microsoft Hololens, Magic Leap One, and nReal are examples of such AR HMDs and provide a relatively low Field of View (FoV) of <60$^{\circ}$. Products like Meta 2 and open source Leap Motion Project North Star both provide wider FoV (90-110$^{\circ}$) but have not yet hit the mass market. We believe the arrival of wide FoV AR HMDs is eminent.

We used AR-SIM (a tool we developed separately) as our development platform, allowing us to use commercially available Virtual Reality (VR) hardware to experiment with experiences (roughly) as they would appear in AR. This aallows us to emulate a VR "real world" as well as the additive-only blending implied by an optical-pass through design with Leap Motion hand tracking or controllers for interaction. After several experiments to determine which sort of effects work well in the optical pass-through AR context, we developed 4 applications and an application manager to examine how well interaction modalities work for users. The application manager uses a forearm-based interaction method. The 4 applications are as follows:

\begin{enumerate}
    \item \textbf{Sticky Notes:} Create custom sized floating/docked canvases, write or draw on them, move them around the space
    \item \textbf{PreviewAR:} An online shopping companion, preview 3D items in physical space by dragging them from your phone
    \item \textbf{Walking Navigator:} An AR navigation system with virtual street signs and ground-based display
    \item \textbf{Control Anything:} An AR smart home system connecting virtual inputs to physical outputs
\end{enumerate}

\section{Application Details}
\begin{figure}
    \centering
    \includegraphics[height=1.2in]{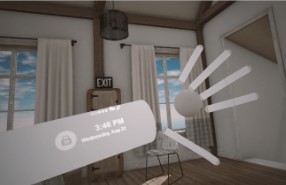}
    \includegraphics[height=1.2in]{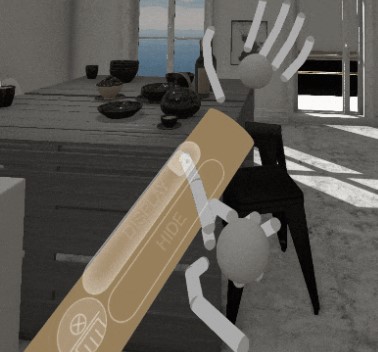}
    \caption{A forearm located application manager}
    \label{fig:appmanager}
\end{figure}

An important aspect of interacting with any application-based device is having the ability to start new applications, switch between running applications and close applications when they are no longer useful. Figure~\ref{fig:appmanager} shows our application manager, which is placed on the (virtual) left forearm of the user and allows the user to slide to unlock functionality on the top of the arm. Once unlocked, the user can launch, show, hide, and close applications as desired on the bottom of the arm.


\subsection{Sticky Notes}
The left-most picture in Figure~\ref{fig:teaser} shows the first application developed, a simple note tool that allows the user to place reminders and drawings in the space around them. A user is able to create a new note by placing their controllers or pinched fingers on the corners of a rectangle then releasing the button or pinch to create the note. An extended finger or controller pointer can be used to draw on the note. A few simple tools were also included to change the size of writing point, clear the note's contents and delete the note entirely. 


\subsection{PreviewAR}
The right-most picture in Figure~\ref{fig:teaser} shows PreviewAR, an online shopping assistant. The application allows a user to browse a catalog of products as they normally would on a phone or tablet. When the user identifies an object of interest, she or he can select/examine a (small) 3D preview of that object just outside the shopping device display, then drag this object into the real world. As the preview is moved away from the shopping device, it expands to 1:1 scale and can be placed where the user likes. This allows users to gauge both spatial and stylistic constraints before buying.

\subsection{Walking Navigator}
\begin{figure}
    \centering
    \includegraphics[height=1.1in]{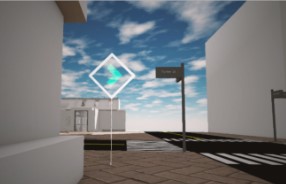}
    \includegraphics[height=1.1in]{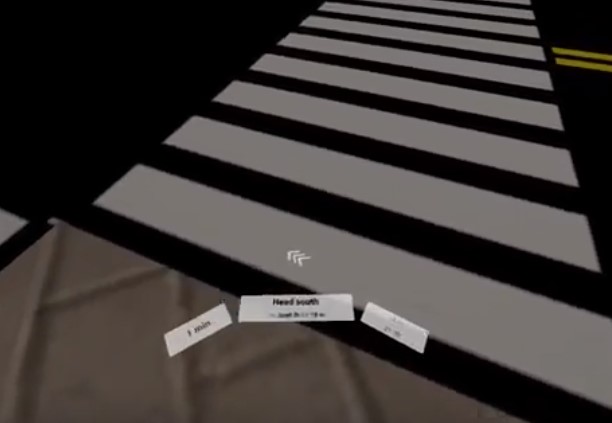}
    \caption{An AR walking navigator application}
    \label{fig:walking}
\end{figure}

Figure~\ref{fig:walking} shows our walking navigation application. This application attempts to balance a natural, clutter-free, safety-first navigation experience with the level of navigation information we have come to expect from our smart devices. A set of minimal, virtual, "street sign"-style navigation markers are coupled with an informative ground-based display that provides remaining time/distance, ETA, and turn-by-turn updates. There's also an optional 2D, hand held mini-map for more detailed exploration. Through this mix we hope to create an experience that can be easily tailored to users with various interests/levels of familiarity with the area.

\subsection{Control Anything}
The explosion of smart home devices with remote connectivity offers an interesting challenge and opportunity for AR devices. Lights, appliances, electronic devices and more can now offer virtual controls and indicators embedded in the real world. The center image of figure~\ref{fig:teaser} shows our control anything application. This application allows the user to create and relocate controls (buttons, switches, and sliders), map them to smart objects (lamp, radio, television), then use them for control of these devices in daily life.

\section{Conclusion}
Through implementing these prototype applications we learned several valuable lessons about designing user experiences for optical pass-through AR. Possibly most importantly, we are more convinced of the significance of developing wider FoV AR HMDs than ever before. At times, even the wide FoV of VR HMDs limited our design, in particular keeping objects near the feet of the user forces the user to look down before the content becomes visible. In addition, development for AR requires careful consideration of real world environmental factors, even when that "real world" is itself a virtual reality. 

\begin{acks}
We acknowledge input and support from many others who contributed to the design of these experiments. This list includes David Luebke, Ziyang Shan, Melinne Hay, Antoine Torossian, Milan Singh, Rachel Albert, Alexander Majercik, and Michael Stengel.
\end{acks}

%
\bibliographystyle{ACM-Reference-Format}
\bibliography{sample-base}

%

\end{document}